\documentclass[journal]{IEEEtran}
\usepackage{enumitem}
\usepackage{epsfig}
\usepackage{graphicx}
\usepackage{amsmath}
\usepackage{amsfonts}
\usepackage{amssymb}
\usepackage{array}
\usepackage{rotating}
\usepackage{epstopdf}
\usepackage{caption}
\usepackage{subfig}
\usepackage{gensymb}
\usepackage{multirow}
\usepackage{sansmath}
\usepackage{tabu}
\usepackage{xspace}
\usepackage{xcolor}

\usepackage[hyphens]{url}
\usepackage{afterpage}
\usepackage{booktabs}
\usepackage{eurosym}

\usepackage{hyperref}
\hypersetup{
pdfauthor={P. Drozdowski, C. Rathgeb, A. Dantcheva, N. Damer, C. Busch},
pdfsubject={Biometrics},
pdftitle={Demographic Bias in Biometrics: A Survey on an Emerging Challenge},
pdfkeywords={Biometrics, bias, bias estimation, bias mitigation, demographics, fairness},
bookmarks=true,
}

\newcommand{\etal}{\textit{et~al}.\@\xspace}
\def\eg{\textit{e.g.}\@\xspace} 
\def\ie{\textit{i.e.}\@\xspace} 
 
\def\cf{\textit{cf.}\@\xspace} 
\def\etc{\textit{etc}.\@\xspace} 
\def\wrt{w.r.t.\@\xspace}

\begin{document}

\title{Demographic Bias in Biometrics: \\ A Survey on an Emerging Challenge}

\author{
\vspace{0.5cm}
    \IEEEauthorblockN{P. Drozdowski\IEEEauthorrefmark{1}, C. Rathgeb\IEEEauthorrefmark{1}, A. Dantcheva\IEEEauthorrefmark{2}, N. Damer\IEEEauthorrefmark{3}\IEEEauthorrefmark{4}, C. Busch\IEEEauthorrefmark{1}} \\
    \IEEEauthorblockA{\IEEEauthorrefmark{1}{da/sec - Biometrics and Internet Security Research Group, Hochschule Darmstadt, Darmstadt, Germany}} \\
    \IEEEauthorblockA{\IEEEauthorrefmark{2}Inria, Sophia Antipolis, France} \\
    \IEEEauthorblockA{\IEEEauthorrefmark{3}Fraunhofer Institute for Computer Graphics Research IGD, Darmstadt, Germany} \\
    \IEEEauthorblockA{\IEEEauthorrefmark{4}Mathematical and Applied Visual Computing, TU Darmstadt,
Darmstadt, Germany}
    \vspace{-0.25cm}
}

\markboth{\textbf{Submitted to IEEE Transactions on Technology and Society}}{\textbf{Submitted to IEEE Transactions on Technology and Society)}}
\maketitle

\begin{abstract}
Systems incorporating biometric technologies have become ubiquitous in personal, commercial, and governmental identity management applications. Both cooperative (\eg access control) and non-cooperative (\eg surveillance and forensics) systems have benefited from biometrics. Such systems rely on the uniqueness of certain biological or behavioural characteristics of human beings, which enable for individuals to be reliably recognised using automated algorithms.

Recently, however, there has been a wave of public and academic concerns regarding the existence of systemic bias in automated decision systems (including biometrics). Most prominently, face recognition algorithms have often been labelled as ``racist'' or ``biased'' by the media, non-governmental organisations, and researchers alike. 

The main contributions of this article are: (1) an overview of the topic of algorithmic bias in the context of biometrics, (2) a comprehensive survey of the existing literature on biometric bias estimation and mitigation, (3) a discussion of the pertinent technical and social matters, and (4) an outline of the remaining challenges and future work items, both from technological and social points of view.
\end{abstract}


\section{Introduction}
\label{sec:introduction}
Artificial intelligence systems increasingly support humans in complex decision-making tasks. Domains of interest include learning, problem solving, classifying, as well as making predictions and risk assessments. Automated algorithms have in many cases already outperformed humans and hence are used to support or replace human operators~\cite{Pasquale-BlackBoxSociety-2015}. Those systems, referred to as ``automated decision systems'', can yield various benefits, \eg increased efficiency and decreased monetary costs. At the same time, a number of ethical and legal concerns have been raised, specifically relating to \emph{transparency, accountability, explainability, and fairness of such systems}~\cite{Osoba-AIBias-Rand-2017}. Automated algorithms can be utilised in diverse critical areas such as criminal justice~\cite{Washington-ProPublicaDebate-2018}, healthcare~\cite{Yu-MedicineBias-2019}, creditworthiness~\cite{Hurley-CreditScoring-2017}, and others~\cite{Castelluccia-AlgorithmicDecisionMaking-2019}, hence often sparking controversial discussions. This article focuses on \emph{algorithmic bias and fairness in biometric systems \wrt demographic attributes}. In this context, an algorithm is considered to be biased if significant differences in its operation can be observed for different demographic groups of individuals (\eg females or dark-skinned people), thereby privileging and disadvantaging certain groups of individuals.

\subsection{Motivation}
\label{subsec:motivation}
The interest and investment into biometric technologies is large and rapidly growing according to various market value studies~\cite{GMI-BiometricMarket-2017, MM-BiometricMarket-2018, Bayometric-BiometricMarket}. Biometrics are utilised widely by governmental and commercial organisations around the world for purposes such as border control, law enforcement and forensic investigations, voter registration for elections, as well as national identity management systems. Currently, the largest biometric system is operated by the Unique Identification Authority of India, whose national ID system (Aadhaar) accommodates almost the entire Indian population of 1,25 billion enrolled subjects at the time of this writing, see the online dashboard~\cite{UIDAI-Dashboard} for live data. 

In recent years, reports of demographically unfair/biased biometric systems have emerged (see section~\ref{sec:biasinbiometrics}), fueling a debate on the use, ethics, and limitations of related technologies between various stakeholders such as the general population, consumer advocates, non-governmental and governmental organisations, academic researchers, and commercial vendors. Such discussions are intense and have even raised demands and considerations that biometric applications should be discontinued in operation, until sufficient privacy protection and demographic bias mitigation can be achieved\footnote{\url{https://www.banfacialrecognition.com/}}\textsuperscript{,}\footnote{\url{https://www.cnet.com/news/facial-recognition-could-be-temporarily-banned-for-law-enforcement-use/}}\textsuperscript{,}\footnote{\url{https://www.theguardian.com/technology/2020/jan/17/eu-eyes-temporary-ban-on-facial-recognition-in-public-places}}\textsuperscript{,}\footnote{\url{https://www.biometricupdate.com/202001/eu-no-longer-considering-facial-recognition-ban-in-public-spaces}}. Algorithmic bias is considered to be one of the important open challenges in biometrics by Ross~\etal~\cite{Ross-OpenProblems-2019}.

\subsection{Article Contribution and Organisation}
\label{subsec:organisation}
In this article, an overview of the emerging challenge of algorithmic bias and fairness in the context of biometric systems is presented. Accordingly, the biometric algorithms which might be susceptible to bias are summarised; furthermore, the existing approaches of bias estimation and bias mitigation are surveyed. The article additionally discusses other pertinent matters, including the potential social impact of bias in biometric systems, as well as the remaining challenges and open issues in this area.

The remainder of this article is organised as follows: relevant background information is provided in section~\ref{sec:background}. Section~\ref{sec:biasinbiometrics} contains a comprehensive survey of the scientific literature on bias estimation and mitigation in biometric systems. Other relevant matters are discussed in section~\ref{sec:discussion}, while concluding remarks and a summary are presented in section~\ref{sec:summary}.

\section{Background}
\label{sec:background}
The following subsections provide relevant background information \wrt the topic of bias in automated decision systems in general (subsection~\ref{subsec:biasintro}) and the basics of biometric systems (subsection~\ref{subsec:biometricsintro}). Furthermore, due to the sensitive nature of the matter at hand, subsection~\ref{subsec:nomenclature} outlines the choices made \wrt the nomenclature used throughout the article.

\subsection{Bias in Automated Decision Systems}
\label{subsec:biasintro}
In recent years, numerous concerns have been raised regarding the accuracy and fairness of automated decision-making systems. For instance, many studies regarding the risk assessment and welfare distribution tools found a number of issues concerning systemic bias and discrimination of the systems' predictions (\eg against dark-skinned people). The impact of such automated decisions on the lives of the affected individuals can be tremendous, \eg being jailed, denied a bail, parole, or welfare payments~\cite{Osoba-AIBias-Rand-2017,Washington-ProPublicaDebate-2018,Garvie-PerpetualLineUp-2016,ONeil-Weapons-2016}. Demographics-based bias and discrimination are especially concerning in this context, even if they occur unintentionally. One would intuitively expect that certain decisions be impacted exclusively by hard facts and evidence, and not factors often associated with discrimination -- such as sex or race, or other context-specific discriminatory factors. Nonetheless, biases in decision-making are a common occurrence; along with notions of fairness, this topic has been extensively studied from the point of view of various disciplines such as psychology, sociology, statistics, and information theory~\cite{Evans-HumanBias-1989,Banks-Bias-2006,Friedman-StatisticsBook-2009}. Recently, the field of bias and fairness in automated computer algorithms and machine learning has emerged~\cite{Zweig-AlgoritmsChances-2018,Mehrabi-BiasSurvey-2019}.

A good discussion of the topic of bias was provided by Danks and London~\cite{Danks-AlgorithmicBias-IJCAI-2017}, as well as Friedman and Nissenbaum~\cite{Friedman-BiasComputers-TOIS-1996}, both of which explored various sources and types of bias in the context of computer systems. In many cases, bias in the automated decision systems is directly related to the human designers or operators of a system. Semi-automatic decision systems are a good example of this. In such systems, a human decision maker can be aided by an algorithm (\eg risk-assessment). In such cases, errors in interpretation of the results of the system might occur; in other words, the human might misunderstand or misrepresent the outputs or general functioning principles of an algorithm~\cite{Lansing-COMPASAssessment-2012, Desmarais-RiskAssessment-2013, Chouldechova-BiasRecidivism-2017}. Furthermore, it has been shown that humans in general tend to over-rely on such automated systems, \ie to overestimate the accuracy of their results~\cite{Mosier-AutomationBias-1998}. While \emph{human} cognitive biases are an important and actively researched topic, this article focuses exclusively on bias occurring in the context of \emph{automated} algorithms themselves. Human congnitve biases have been analysed \eg by Evans~\cite{Evans-HumanBias-1989}, whereas bias in human interactions with automated system was explored \eg by Parasuraman and Manzey~\cite{Parasuraman-AutomationBias-2010}.

In the context of automated decision algorithms themselves, numerous potential bias causes exist. Most prominently, the \emph{training data} could be skewed, incomplete, outdated, disproportionate or have embedded historical biases, all of which are detrimental to algorithm training and propagate the biases present in the data. Likewise, the \emph{implementation} of an algorithm itself could be statistically biased or otherwise flawed in some way, for example due to moral or legal norms, poor design, or data processing steps such as parameter regularisation or smoothing. For more details on the topic of algorithmic bias in general, the reader is referred to \eg~\cite{Castelluccia-AlgorithmicDecisionMaking-2019,Danks-AlgorithmicBias-IJCAI-2017,Friedman-BiasComputers-TOIS-1996}. In the next sections, an introduction to biometric systems is provided, followed by a survey on algorithmic bias in such systems specifically.

\subsection{Biometric Systems}
\label{subsec:biometricsintro}
Biometric systems aim at establishing or verifying the identity or demographic attributes of individuals. In the international standard ISO/IEC 2382-37~\cite{ISO-Vocabulary-2017}, ``biometrics'' is defined as: ``automated recognition of individuals based on their biological and behavioural characteristics''.

\begin{figure}[!ht]
\fboxsep=0mm
\fboxrule=1.5pt
\centering
\subfloat[Face]{\includegraphics[height=2.7cm]{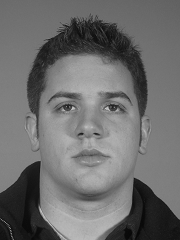}} \hfill
\subfloat[Iris]{\includegraphics[height=2.7cm]{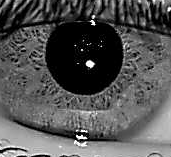}} \hfill
\subfloat[Fingerprint]{\includegraphics[height=2.7cm]{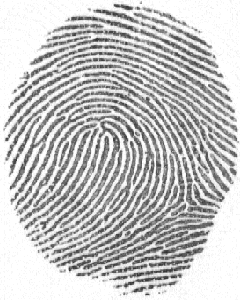}} \hfill
\subfloat[Veins]{\includegraphics[height=2.7cm]{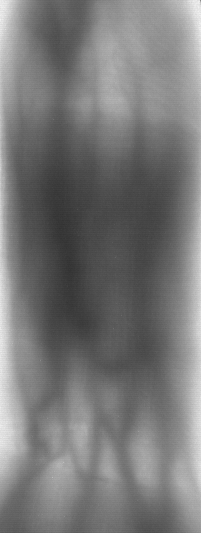}}
\caption{Examples of biometric characteristics (images from publicly available research databases~\cite{Phillips-FRGC-CVPR-2005,Kumar-IITD-2010,Ortega-MCYT-2003,Ton-UTFVP-2013}).}
\label{fig:characteristicsexamples}
\end{figure}

Humans possess, nearly universally, physiological characteristics which are highly distinctive and can therefore be used to distinguish between different individuals with a high degree of confidence. Example images of several prominent biometric characteristics are shown in figure~\ref{fig:characteristicsexamples}.

Broadly speaking, an automated biometric system consists of: (1) a capture device (\eg a camera), with which the biometric samples (\eg images) are acquired; (2) a database which stores the biometric information and other personal data; (3) signal processing algorithms, which estimate the quality of the acquired sample, find the region of interest (\eg a face), and extract the distinguishing features from it; (4) comparison and decision algorithms, which enable ascertaining of similarity of two biometric samples by comparing the extracted feature vectors and establishing whether or not the two biometric samples belong to the same source.

In the past, biometric systems typically utilised handcrafted features and algorithms (\ie texture descriptors, see Liu~\etal~\cite{Liu-TextureSurvey-2019}). Nowadays, the use of machine learning and deep learning has become increasingly popular and successful. Relevant related works include~\cite{Taigman-DeepFace-CVPR-2014,Schroff-Facenet-CVPR-2015,Parkhi-VGGFace-BMVC-2015}, which achieved breakthrough biometric performances in facial recognition. Furthermore, promising results for deep learning-based fingerprint (see \eg~\cite{Tang-FingerNet-2017}) and iris (see \eg~\cite{Nguyen-IrisNet-2018}) recognition have also been achieved. For a review of deep learning techniques applied within biometrics, the reader is referred to Sundararajan and Woodard~\cite{Sundararajan-DeepLearningBiometrics-2018}. For a highly comprehensive introduction to biometrics, the reader is referred to Li and Jain~\cite{Li-BiometricsEncyclopedia-2015} and the handbook series~\cite{Jain-HandbookBiometrics-2007,Li-HandbookFace-2004,Maltoni-HandbookFingerprint-2009,Bowyer-HandbookIris-2016,Uhl-HandbookVascular-2020}.

\subsection{Nomenclature}
\label{subsec:nomenclature}
In this section, the nomenclature used throughout this article is explained. The authors note that demographic words, groups, and concepts such as ``gender'', ``sex'', ``race'', and ``ethnicity'' can be extremely divisive and bear a heavy historical, cultural, social, political, or legislative load. The authors do not seek to define or redefine those terms; we merely report on the current state of the research. In the literature surveyed later on in this article, following trends can be distinguished:

\begin{enumerate}
\item The terms ``gender'' and ``sex'' are often used in a binary and conflated manner. Readers interested in the possible consequences of this narrow approach are referred to~\cite{Keyes-MisgenderingMachines-2018}.
\item Similarly, very often no real distinction between the terms ``race'' and ``ethnicity'' is made; moreover, the typical categorisation is very coarse, only allowing for a small and finite (less than ten) possible racial/ethnic categories.
\item In general, and especially in the case of facial biometrics, demographic factors seem to be considered on the phenotypic basis, \ie concerning the observable traits of the subjects (\eg colour of the skin or masculine appearance).
\end{enumerate}

Due to the demographic terms carrying a large amount of complexity and potential social divisiveness, the authors do not engage in those debates in this article, and merely reproduce and discuss the technical aspects of the current research. For the sake of consistency, certain decisions regarding the used nomenclature have to be made, especially since the surveyed literature does often seem to use the aforementioned demographic terms ambiguously or interchangeably.

Recently, in the context of biometrics, ISO/IEC has made the following separation~\cite{ISO-Bias}\footnote{Note that the document is currently in a draft stage.}: while the term ``gender'' is defined as ``the state of being male or female as it relates to social, cultural or behavioural factors'', the term ``sex'' is understood as ``the state of being male or female as it relates to biological factors such as DNA, anatomy, and physiology''. The report also defines the term ``ethnicity'' as ``the state of belonging to a group with a common origin, set of customs or traditions'', while the term ``race'' is not defined there. While the cultural and religious norms can certainly affect biometric operations, the surveyed literature mostly considers the appearance-based features and categorisation -- hence, the term ``race'' is used instead of ``ethnicity'' and the term ``sex'' is used instead of ``gender'' in accordance with ISO/IEC 22116~\cite{ISO-Bias}. In the context of biometrics in general, the standardised biometric vocabulary is used, see ISO/IEC 2382-37~\cite{ISO-Vocabulary-2017}. Finally, it is noted that a large part of the surveyed biometric literature follows the notions and metrics regarding evaluation of biometric algorithms irrespective of the chosen biometric characteristic defined in ISO/IEC 19795-1~\cite{ISO-PerformanceReporting-2006}.

Those limitations and imprecisions of the nomenclature notwithstanding, due to the potential of real and disparate impacts~\cite{Feldman-DisparateImpact-ICKDDM-2015} of automated decision systems including biometrics, it is imperative to study the bias and fairness of such algorithms \wrt the demographic attributes of the population, regardless of their precise definitions.

\begin{figure*}[!ht]
\centering
\subfloat[Verification.]{\includegraphics[width=\columnwidth]{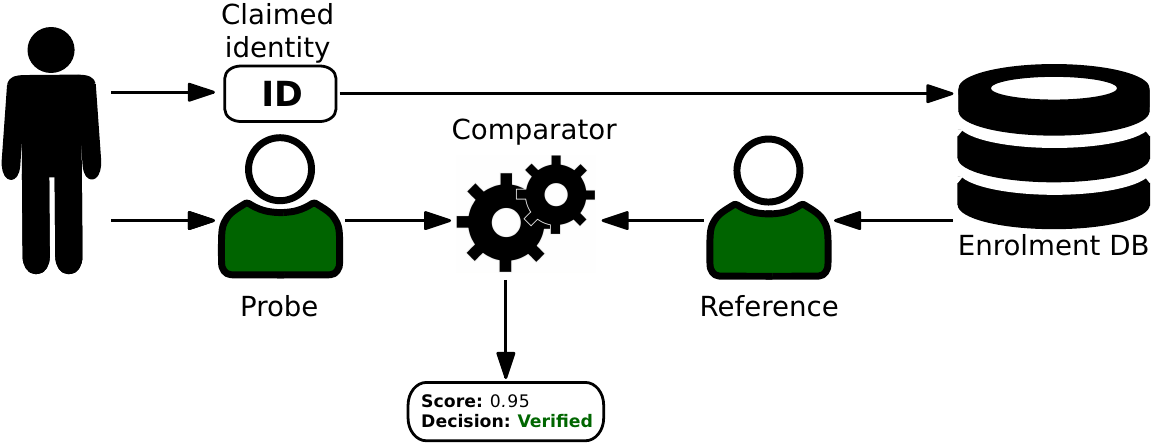}} \hfill
\subfloat[Identification.]{\includegraphics[width=\columnwidth]{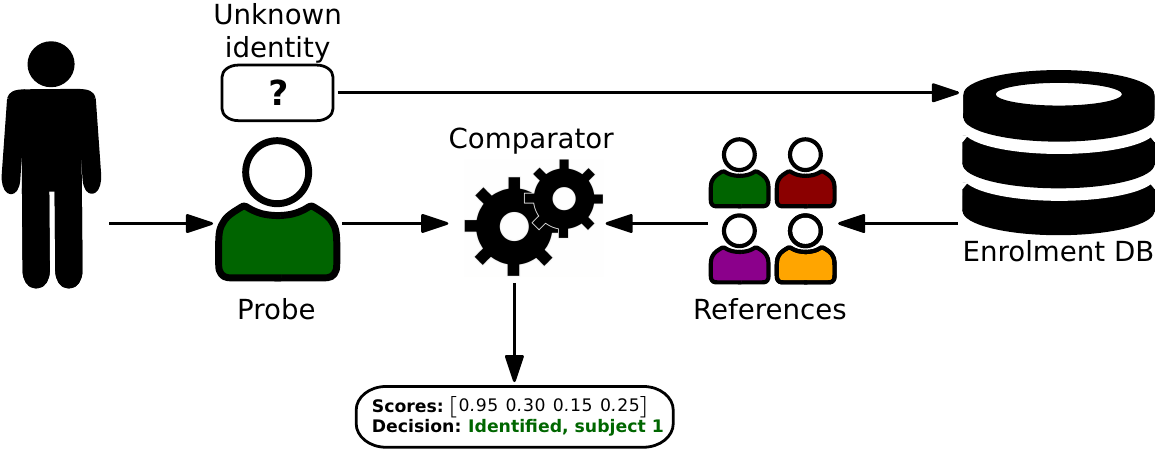}}
\\
\subfloat[Classification and estimation.]{\includegraphics[height=2cm]{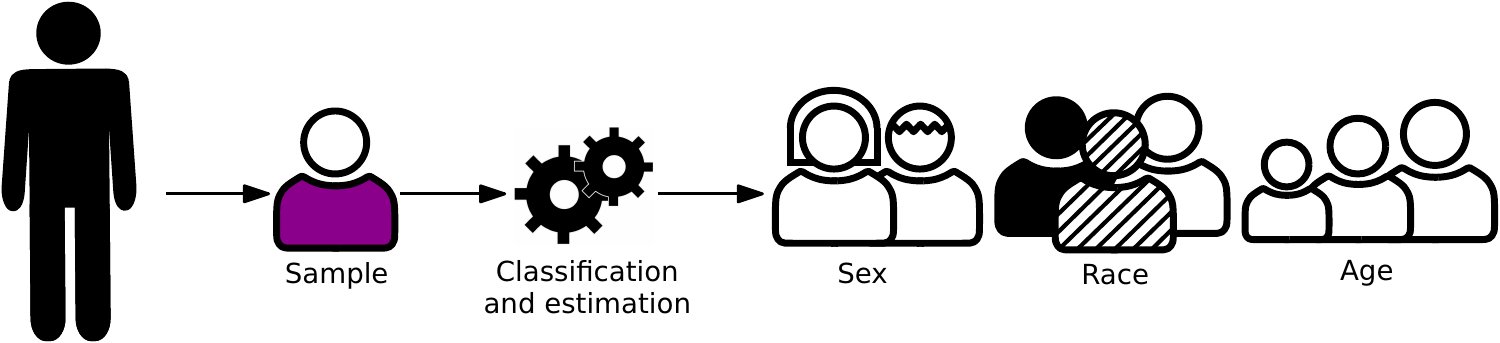}} \hfil
\subfloat[Quality assessment.]{\includegraphics[height=2cm]{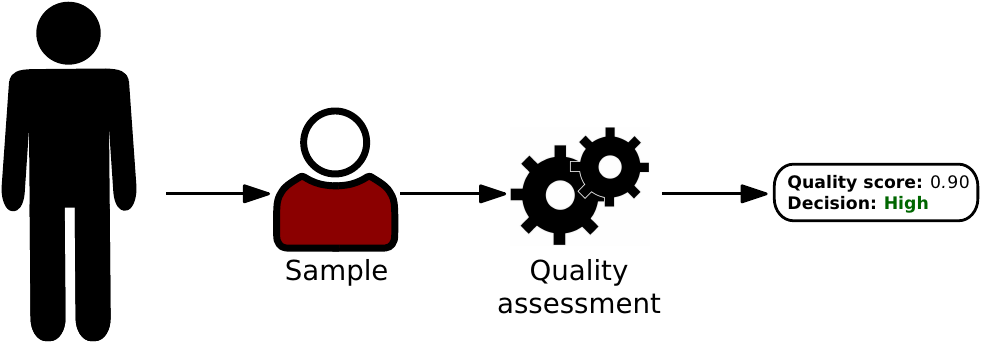}} \\
\subfloat[Segmentation and feature extraction.]{\includegraphics[height=2cm]{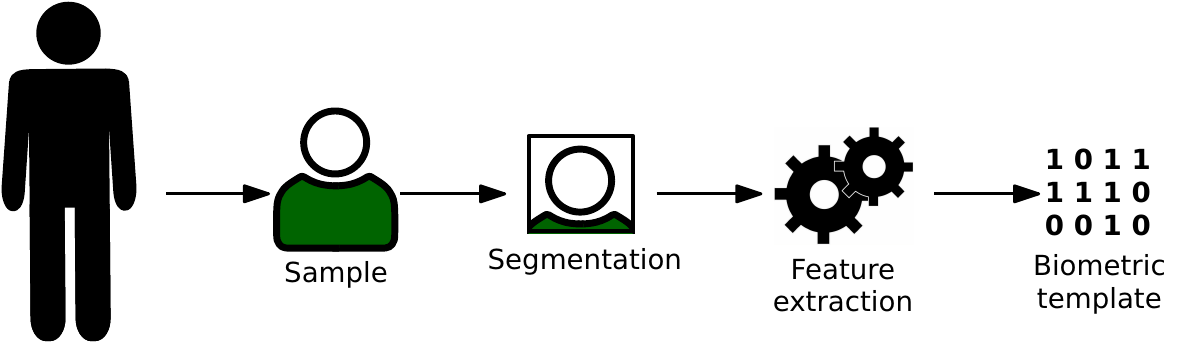}} \hfil
\subfloat[Presentation attack detection.]{\includegraphics[height=2cm]{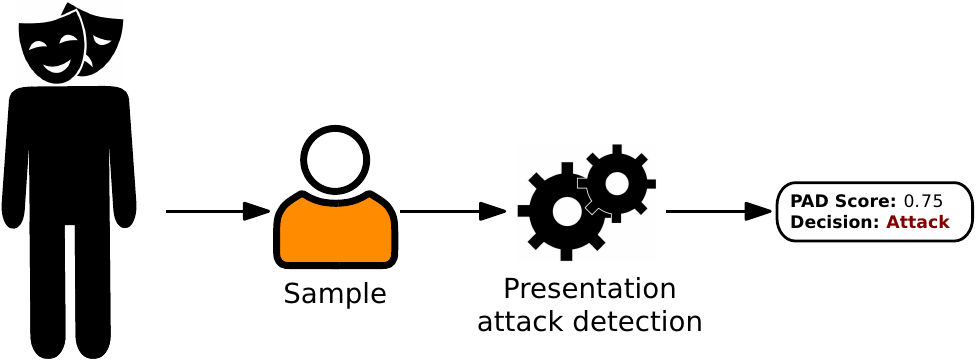}}
\caption{Conceptual overview of algorithms used in biometric systems.} 
\label{fig:algorithms}
\end{figure*}

\section{Bias in Biometric Systems}
\label{sec:biasinbiometrics}
To facilitate discussions on algorithmic fairness in biometric systems, Howard~\etal~\cite{Howard-DemographicEffectsFace-2019} introduced following two terms:

\begin{description}
\item[Differential performance] concerns the differences in (genuine and/or impostor) score distributions between the demographic groups. Those effects are closely related to the so-called ``biometric menagerie''~\cite{Doddington-SheepGoatsLambsWolves-1998,Yager-Menagerie-2007,Daugman-Doppelgangers-2016}. While the menagerie describes the score distributions being statistically different for specific individual subjects, the introduced term describes the analogous effect for different demographic groups of subjects.
\item[Differential outcomes] relate to the decision results of the biometric system, \ie the differences in the false-match and false-non-match rates at a specific decision threshold.
\end{description}

Given that these terms have been introduced relatively recently, the vast majority of surveyed literature has not (directly) used them, instead \textit{ad hoc} methodologies based on existing metrics were used. However, Grother~\etal~\cite{Grother-NIST-FRVTBias-2019} presented a highly comprehensive study of the demographic effects in biometric recognition, conducting their benchmark utilising the terms and notions above. A standardisation effort in this area under the auscpices of ISO/IEC is ongoing~\cite{ISO-Bias}.

Before surveying the literature on bias estimation and mitigation (subsections~\ref{subsec:estimation} and~\ref{subsec:mitigation}, respectively), this section begins with an outline of biometric algorithms which might be affected by bias (subsection~\ref{subsec:algorithms}), as well as of covariates which might affect them (subsection~\ref{subsec:covariates}).

\subsection{Algorithms}
\label{subsec:algorithms}
Similarly to other automated decision systems, human biases have been shown to exist in the context of biometrics. The so-called ``other-race effect'' has long been known to affect human ability to recognise faces~\cite{Furl-OtherRaceEffect-2002}. As previously stated, the cognitive biases of humans are out of scope for this article, as it focuses on the biases in the algorithms themselves. The processing pipeline of a biometric system can consist of various algorithms depending on the application scenario and the chosen biometric characteristic. Said algorithms might be subject to algorithmic bias \wrt certain covariates, which are described in subsection~\ref{subsec:covariates}. Below, the most important algorithms used in the context of biometrics are described and visualised conceptually in figure~\ref{fig:algorithms}.

One of the most prevalent uses of biometrics is recognition. Here, distinguishing features of biometric samples are compared to ascertain their similarity. Such systems typically seek to (1) determine if an individual is who they claim to be (\ie one-to-one comparison), or (2) to determine the identity of an individual by searching a database (\ie one-to-many search). \newpage Accordingly, the following two scenarios might be used in biometric recognition: 

\begin{description}
\item[Verification] Referring to the ``process of confirming a biometric claim through biometric comparison''~\cite{ISO-Vocabulary-2017,ISO-PerformanceReporting-2006}.
\item[Identification] Referring to the ``process of searching against a biometric enrolment database to find and return the biometric reference identifier(s) attributable to a single individual''~\cite{ISO-Vocabulary-2017,ISO-PerformanceReporting-2006}.
\end{description}

The biometric samples are a rich source of information beyond the mere identity of the data subject. Another use case of biometrics is the extraction of auxiliary information from a biometric sample, primarily using the following algorithms: 

\begin{description}
\item[Classification and estimation] Referring to the process of assigning demographic or other labels to biometric samples~\cite{Dantcheva-SoftBiometricsSurvey-TIFS-2016,Sun-SoftBiometrics-2018}.
\end{description}

Prior to recognition or classification tasks, the system must acquire and pre-process the biometric sample(s). Here, most prominently, following algorithms might be used:

\begin{description}
\item[Segmentation and feature extraction] Referring to the process of locating the region of interest and extracting a set of biometric features from a biometric sample~\cite{Li-BiometricsEncyclopedia-2015}.
\item[Quality assessment] Referring to the process of quantifying the quality of an acquired biometric sample~\cite{ISO-Quality-2016,Bharadwaj-SampleQuality-2014}.
\item[Presentation attack detection (PAD)] Referring to the ``automated determination of a presentation attack'', \ie detecting a ``presentation to the biometric data capture subsystem with the goal of interfering with the operation of the biometric system''~\cite{ISO-PAD-2016,Marcel-AntiSpoofing-2019}.
\end{description}

\begin{figure*}[!ht]
\fboxsep=0mm
\fboxrule=1.5pt
\centering
\subfloat[Demographic (different sex, age, and race).]{\includegraphics[width=0.3\textwidth]{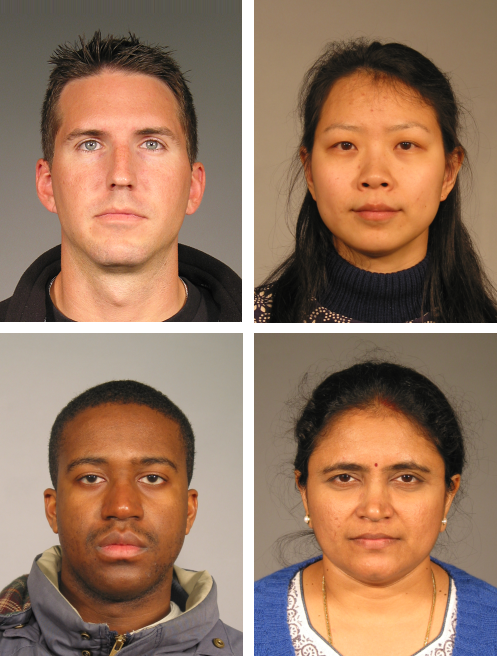}} \hspace{0.2cm}
\subfloat[Subject-specific (different pose and expression, use of make-up and accessories).]{\includegraphics[width=0.3\textwidth]{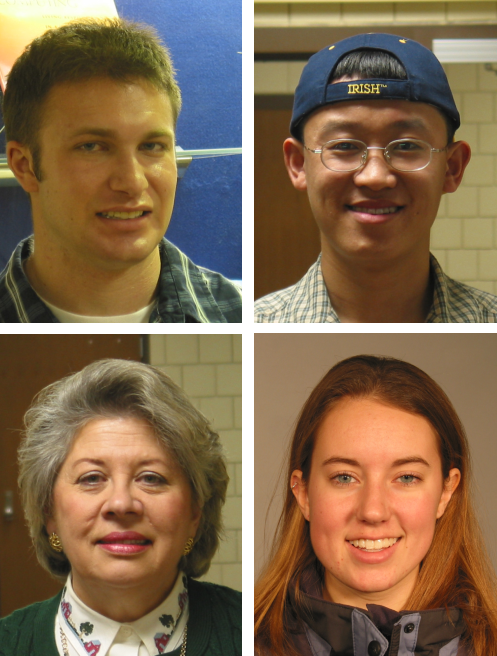}} \hspace{0.2cm}
\subfloat[Environmental (different lighting conditions, sharpness, and resolution).]{\includegraphics[width=0.3\textwidth]{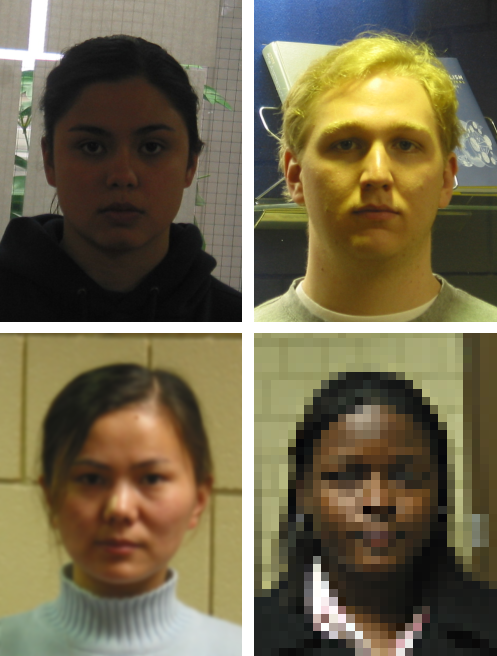}}
\caption{Example images of covariates which might influence a biometric system utilising facial information (images from a publicly available research database~\cite{Phillips-FRGC-CVPR-2005}).}
\label{fig:covariates}
\end{figure*}

\subsection{Covariates}
\label{subsec:covariates}
Broadly, three categories of covariates relevant for the effectiveness of the biometric algorithms can be distinguished:

\begin{description}
\item[Demographic] Referring to \eg the sex, age, and race of the data subject.
\item[Subject-specific] Referring to the behaviour of the subject (\eg pose or expression, use of accessories such as eyewear or make-up), as well as their interaction with the capture device (\eg distance from a camera or pressure applied to a touch-based sensor).
\item[Environmental] Referring to the effects of the surroundings on the data acquisition process (\eg illumination, occlusion, resolution of the images captured by the sensor).
\end{description}

Figure~\ref{fig:covariates} shows example images of the aforementioned covariates using the facial biometric characteristic. While there do exist studies that investigate environmental and subject-specific covariates (\eg~\cite{Kortylewski-FaceDatasetBias-2018}), this article concentrates on the demographic covariates.

\begin{table*}[!ht]
\centering
\caption{Summary of studies concerning bias estimation in biometric systems.}
\label{table:estimationstudies}
\resizebox{\textwidth}{!}{
\begin{tabular}{llp{3.25cm}lp{11.75cm}}
\toprule
\textbf{Reference} & \textbf{Characteristic} & \textbf{Algorithm(s)} & \textbf{Covariate(s)} &  \textbf{Key Findings} \\
\midrule
Beveridge~\etal~\cite{Beveridge-FaceBiasFRVT-2009} & Face & Verification & Sex, age, race & Better biometric performance for older subjects, males, and East Asians. \\
Lui~\etal~\cite{Lui-FaceBiasMeta-2009} & Face & Verification & Sex, age, race & Meta-analysis of previous studies. \\
Guo~\etal~\cite{Guo-AgeClassification-2010} & Face & Age estimation & Sex, race & Large impact of the training data composition on the system accuracy. \\
Grother~\etal~\cite{Grother-FaceRec2d-2010} & Face & Verification & Sex & More false-non-matches at fixed FMR for females than for males. \\
Phillips~\etal~\cite{Phillips-FaceRace-2011} & Face & Verification & Race & Varying results depending on the demographic origin of the algorithm and demographic structure of the data subjects. \\
O{'}Toole~\etal~\cite{Otoole-FaceDemographicEffects-2012} & Face & Verification & Sex, race & The concept of ``yoking'' in experimental evaluation to  demonstrate the variability of algorithm performance estimates. \\
Klare~\etal~\cite{Klare-FaceDemographics-2012} & Face & Verification & Sex, age, race & Lower biometric performance for females, young, and black cohorts. \\
Givens~\etal~\cite{Givens-FaceRecognitionPerformance-2013} & Face & Verification & Sex, age, race & Better biometric performance for Asian and older subjects. \\
Beveridge~\etal~\cite{Beveridge-FG-2015} & Face & Verification & Sex, race & Better biometric performance for males and Asian subjects. \\
Ricanek~\etal~\cite{Ricanek-FaceChildren-2015} & Face & Verification & Age & Poor biometric performance for children. \\
El Khiyari~\etal~\cite{Elkhiyari-FaceBias-2016} & Face & Verification & Sex, age, race & Lower biometric performance for female, 18-30 age group, and dark-skinned subjects. \\
Deb~\etal~\cite{Deb-FaceAging-2017} & Face & Verification & Sex, race & Algorithm dependent effects of the covariates. \\
Best-Rowden~\etal~\cite{BestRowden-FaceAgeing-2018} & Face & Verification & Sex, age, race & Lower comparison scores for females. \\
Buolamwini~\etal~\cite{Buolamwini-GenderShades-2018} & Face & Sex and race  classification & Race & Highest accuracy for males and light-skinned individuals; worst accuracy for dark-skinned females. \\
Deb~\etal~\cite{Deb-FaceChildren-2018} & Face & Verification, identification & Age & Child females easier to recognise than child males. \\
Michalski~\etal~\cite{Michalski-FaceAge-2018} & Face & Verification & Age & Large variation of biometric performance across age and ageing factors in children. Poor biometric performance for very young subjects. \\
Abdurrahim~\etal~\cite{Abdurrahim-FaceBiasReview-2018} & Face & Verification & Sex, age, race & Lower biometric performance for females, inconsistent results \wrt age and race. \\
Rhue~\etal~\cite{Rhue-RacialEmotion-2018} & Face & Emotion classification & Race & Negative emotions more likely to be assigned to dark-skinned males. \\
Lu~\etal~\cite{Lu-FaceVerificationBias-2019} & Face & Verification & Sex, age, race & Lower biometric performance for females; better biometric performance for middle-aged. \\
Raji~\etal~\cite{Raji-ActionableAuditing-2019} & Face & Sex and race classification & Sex, race & Lower accuracy for dark-skinned females. \\
Srinivas~\etal~\cite{Srinivas-FaceGenderAgeBias-2019} & Face & Verification, identification & Sex, age & Lower biometric performance for females and children. \\
Cook~\etal~\cite{Cook-DemographicEffectsFace-2019} & Face & Verification & Sex, age, race & Genuine scores tend to be worse for females than males. \\
Hupont~\etal~\cite{Hupont-BiasFace-2019} & Face & Verification & Sex, race & Highest biometric performance for white males, lowest for Asian females. \\
Denton~\etal~\cite{Denton-BiasDetection-2019} & Face & Classification & CelebA attributes & Generative adversarial model which can reveal biases in a face attribute classifier. \\
Garcia~\etal~\cite{Garcia-FaceBias-2019} & Face & Verification, presentation attack detection & Sex, race & Higher inter-subject distance for Caucasians than other groups; morphing attacks much more successful for Asian females. \\
Nagpal~\etal~\cite{Nagpal-DLPrideorPrejudice-2019} & Face & Verification & Age, race & Training data dependent own-age and own-race effect in DNN-based systems. \\
Krishnapriya~\etal~\cite{Krishnapriya-FaceRaceVariability-2019} & Face & Quality, verification & Race & Lower rate of ICAO compliance~\cite{ISO-FaceDataFormat-2005} for the dark-skinned cohort, fixed decision thresholds not suitable for cross-cohort biometric performance benchmark. \\
Muthukumar~\cite{Muthukumar-GenderClassificationBias-2019} & Face & Sex classification & Race & Lower accuracy for dark females; importance of not only skin type, but also luminance in the images on the results. \\
Srinivas~\etal~\cite{Srinivas-BiasAge-2019} & Face & Verification, identification & Age & Lower biometric performance for children. \\
Vera-Rodriguez~\etal~\cite{Vera-facegenderid-2019} & Face & Verification & Sex & Lower biometric performance for females. \\
Howard~\etal~\cite{Howard-DemographicEffectsFace-2019} & Face & Verification & Sex, age, race & Evaluates effects of population homogeneity on biometric performance. \\
Wang~\etal~\cite{Wang-RacialFacesInTheWild-2019} & Face & Verification & Race & Higher biometric performance for Caucasians \\
Serna~\etal~\cite{Serna-DeepLearningFaceBias-2019} & Face & Verification & Sex, race & Better biometric performance for male Caucasians; large impact of the training data composition on the system accuracy. \\ Cavazos~\etal~\cite{Cavazos-FaceRace-2019} & Face & Verification & Sex, race & Higher false match rate for Asians compared to Caucasians at operationally relevant fixed decision thresholds; data-driven anomalies might contribute to system bias. \\
Grother~\etal~\cite{Grother-NIST-FRVTBias-2019} & Face & Verification, identification & Sex, age, race & Large-scale benchmark of commercial algorithms. Algorithm dependent false positive differentials \wrt race. Consistently elevated false positives for female, elderly and very young subjects. Algorithm specific false negative differentials, also correlated with image quality. \\
Robinson~\etal~\cite{Robinson-FaceBias-2020} & Face & Verification & Sex, race & Highest biometric performance for males and Caucasians. \\
Albiero~\etal~\cite{Albiero-FaceGender-2020} & Face & Verification & Sex & Lower biometric performance for females. Negative impact of facial cosmetics on (female) genuine scores distribution. Minor impact of expression, pose, hair occlusion, and imbalanced datasets on bias. \\
Krishnapriya~\etal~\cite{Krishnapriya-FaceBiasRace-2020} & Face & Verification, identification & Race & Lower biometric performance for females, higher false match rate for African-Americans, and higher false non-match rate for Caucasians at fixed, operationally relevant decision threshold. \\
Terh{\"o}rst~\etal~\cite{Terhorst-BiasFaceQuality-2020} & Face & Quality & Age, race & Bias in quality scores for demographic and non-demographic characteristics is significant. Bias is transferred from face recognition to face image quality. \\
\midrule
Hicklin~\etal~\cite{Hicklin-IAFIS-2002} & Fingerprint & Quality & Sex & Lower sample quality for females. \\
Sickler~\etal~\cite{Sickler-FingerprintAge-2005} & Fingerprint & Quality & Age & Lower sample quality for the elderly. \\
Modi~\etal~\cite{Modi-FingerprintAge-2006} & Fingerprint & Quality, verification & Age & Lower sample quality and biometric performance for the elderly. \\
Modi~\etal~\cite{Modi-FingerprintAge-2007} & Fingerprint & Quality, verification & Age & Lower sample quality and biometric performance for the elderly. \\
Frick~\etal~\cite{Frick-FingerprintGenderImpact-2008} & Fingerprint & Quality, verification & Sex & Higher sample quality and biometric performance for males. \\
O{'}Connor~\etal~\cite{Oconnor-FingerprintGender-2011} & Fingerprint & Quality, verification & Sex & Higher sample quality for males, higher biometric performance for females. \\
Schumacher~\etal~\cite{Schumacher-FingerprintAge-2013} & Fingerprint & Quality, verification & Age & Lower sample quality and biometric performance for children. \\
Yoon~\etal~\cite{Yoon-LongitudinalFingerprint-2015} & Fingerprint & Quality, verification & Sex, age, race & Negligible correlations between sample quality and subject age; sex and race have a marginal impact on comparison scores, whereas subject's age has a non-trivial impact for genuine scores. \\
Galbally~\etal~\cite{Galbally-FingerprintAge-2018,Galbally-FingerprintAge-2019} & Fingerprint & Quality, verification & Age & On average, low quality for children under 4 years and elderly (70+ years), medium quality for children between 4 and 12 years. Lowest biometric performance in youngest children, then elderly. \\
Preciozzi~\etal~\cite{Preciozzi-FingerprintAge-2020} & Fingerprint & Quality, verification & Age & Lower sample quality and biometric performance for young children. \\
\midrule
Drozdowski~\etal~\cite{Drozdowski-BiasFingervein-2020} & Fingervein & Verification & Sex, age & No statistically significant biases detected. \\ 
\midrule
Fang~\etal~\cite{Fang-IrisPADBias-2020} & Iris & Presentation attack detection & Sex & Better PAD rates for males. Maps differential performance/outcome concepts to PAD. \\
\midrule
Xie~\etal~\cite{Xie-PalmprintClassification-2018} & Palmprint & Sex classification & Sex & Higher accuracy for females. \\
Uhl~\etal~\cite{Uhl-HandBiometricsChildren-2009} & Palmprint & Verification & Age & Lower biometric performance for very young subjects. \\
\midrule
Brand{\~a}o~\etal~\cite{Brandao-PedestrianDetectionBias-2019} & Unconstrained & Pedestrian detection & Sex, age & Higher miss rate for children. \\
\bottomrule
\end{tabular}
}
\end{table*}

\begin{table*}[!ht]
\centering
\caption{Summary of studies concerning bias mitigation in biometric systems.}
\label{table:mitigationstudies}
\resizebox{\textwidth}{!}{
\begin{tabular}{llll}
\toprule
\textbf{Reference} & \textbf{Characteristic} & \textbf{Algorithm(s)} & \textbf{Method(s)} \\
\midrule
Guo~\etal~\cite{Guo-AgeClassification-2010} & Face & Age classification & Dynamic classifier selection based on the demographic attributes. \\
Klare~\etal~\cite{Klare-FaceDemographics-2012} & Face & Verification, identification & Balanced training dataset or dynamic matcher selection based on the demographic attributes. \\
Guo~\etal~\cite{Guo-FaceUnderrepresentedClasses-2017} & Face & Verification, identification & Imbalanced learning. \\
Ryu~\etal~\cite{Ryu-InclusiveFacenet-2017} & Face & Sex and race classification & Twofold transfer learning, balanced training dataset. \\
Hasnat~\etal~\cite{Hasnat-IdemiaFace-2017} & Face & Verification & Imbalanced learning. \\
Deb~\etal~\cite{Deb-FaceChildren-2018} & Face & Verification, identification & Training fine-tuning. \\
Michalski~\etal~\cite{Michalski-FaceAge-2018} & Face & Verification & Dynamic decision threshold selection. \\
Alvi~\etal~\cite{Alvi-BiasRemoval-2018} & Face & Sex, age, and race classification & Bias removal from DNN embeddings. \\
Das~\etal~\cite{Das-MitigatingClassificationBias-2018} & Face & Sex, age, and race classification & Multi-task CNN with dynamic joint loss. \\
Acien~\etal~\cite{Acien-ClassificationBias-2018} & Face & Verification, identification & Suppression of DNN features related to sex and race. \\
Amini~\etal~\cite{Amini-BiasMitigating-2019} & Face & Detection & Unsupervised learning, sampling probabilities adjustment. \\
Lu~\etal~\cite{Lu-FaceVerificationBias-2019} & Face & Verification & Curating training data (noisy label removal) using automatic sex estimation and clustering. \\
Terh{\"o}rst~\etal~\cite{Terhorst-AttributeSuppression-2019,Terhorst-FacePrivacy-2019} & Face & Sex and age classification & Suppression of demographic attributes. \\
Gong~\etal~\cite{Gong-FaceDebiasing-2019} & Face & Verification; sex, age, and race classification & Disentangled representation for identity, sex, age, and race reduces bias for all estimations. \\
Kortylewski~\etal~\cite{Kortylewski-BiasSyntheticData-2019} & Face & Verification & Synthetic data use in algorithm training. \\
Krishnapriya~\etal~\cite{Krishnapriya-FaceRaceVariability-2019} & Face & Verification & Cohort-dependent decision thresholds. \\
Srinivas~\etal~\cite{Srinivas-BiasAge-2019} & Face & Verification & Score-level fusion of algorithms. \\
Vera-Rodriguez~\etal~\cite{Vera-facegenderid-2019} & Face & Verification & Covariate-specific or covariate-balanced training. \\
Wang~\etal~\cite{Wang-BiasSkewnessLearning-2019} & Face & Verification & Reinforcement learning, balanced training datasets. \\
Robinson~\etal~\cite{Robinson-FaceBias-2020} & Face & Verification, identification & Learning subgroup-specific thresholds mitigate the bias and boost overall performance. \\
Bruveris~\etal~\cite{Bruveris-FaceGeographic-2020} & Face & Verification & Weighted sampling and fine-grained labels. \\
Smith~\etal~\cite{Smith-FaceBiasAugmentation-2020} & Face & Sex and age classification & Data augmentation for model training. \\
Terh{\"o}rst~\etal~\cite{Terhorst-BiasMitigation-2020} & Face & Verification & Individual fairness through fair score normalisation. \\
Terh{\"o}rst~\etal~\cite{Terhorst-FaceBiasMitigation-2020} & Face & Verification, identification & Comparison-level bias-mitigation by learning a fairness-driven similarity function. \\
\midrule
Gottschlich~\etal~\cite{Gottschlich-FingeprintAge-2011} & Fingerprint & Verification, identification & Modelling fingerprint growth and rescaling. \\
Preciozzi~\etal~\cite{Preciozzi-FingerprintAge-2020} & Fingerprint & Quality, verification & Rescaling and bi-cubic interpolation as preprocessing. \\
\midrule
Bekele~\etal~\cite{Bekele-SoftBiometrics-2017} & Unconstrained & Soft-biometric classification & Weighing to compensate for biases from imbalanced training dataset. \\
Wang~\etal~\cite{Wang-BalancedDatasetsNotEnough-2019} & Unconstrained & Classification & Introduces concepts of dataset and model leakage; adversarial debiasing network. \\
\bottomrule

\end{tabular}
}
\end{table*}

\subsection{Estimation}
\label{subsec:estimation}
Table~\ref{table:estimationstudies} summarises the existing research in the area of bias estimation in biometrics. The table is organised conceptually as follows: the studies are divided by biometric characteristic and listed chronologically. The third column lists the algorithms (recall subsection~\ref{subsec:algorithms}) evaluated by the studies, while the covariates (recall subsection~\ref{subsec:covariates}) considered in the studies are listed in the next column. Finally, the last column outlines the key finding(s) of the studies. Wherever possible, those were extracted directly from the abstract or summary sections of the respective studies.

By surveying the existing literature, following trends can be distinguished:

\begin{enumerate}
\item Most existing studies conducted the experiments using \emph{face}-based biometrics. There are significantly fewer studies on other modalities (primarily fingerprint).
\item The majority of studies concentrated on biometric \emph{recognition} algorithms (primarily verification), followed by quality assessment and classification algorithms.
\item Some scenarios have barely been investigated, \eg presentation attack detection.
\item The existing studies predominantly considered the \emph{sex covariate}; the race covariate is also often addressed (possibly due to the recent press coverage~\cite{Orcutt-FaceBias-2016,Snow-ACLU-2018}). The age covariate is least often considered in the context of bias in the surveyed literature. The impact of ageing on biometric recognition is an active field of research, but out of scope for this article. The interested reader is referred to \eg~\cite{BestRowden-FaceAgeing-2018,Yoon-LongitudinalFingerprint-2015,Fairhurst-BiometricAge-2013,Bowyer-IrisAgeing-2015,Grother-IrisAgeing-2015,Arnold-Fignerprint-2005}.
\item Many studies focused on general \emph{accuracy} rather than distinguishing between false positive and false negative errors. Recent works~\cite{Howard-DemographicEffectsFace-2019,Grother-NIST-FRVTBias-2019} introduced and used the useful concepts of ``false positive differentials'' and ``false negative differentials'' in demographic bias benchmarks. 
\item A significant number of studies (\eg~\cite{Howard-DemographicEffectsFace-2019,Grother-NIST-FRVTBias-2019,Cook-DemographicEffectsFace-2019}) conducted evaluations on \emph{sequestered databases and/or commercial systems}. Especially the results of Grother~\etal~\cite{Grother-NIST-FRVTBias-2019} in the context of an evaluation conducted by the National Institute of Standards and Technology (NIST) were valuable due to the realistic/operational nature of the data, the large scale of used databases, as well as the testing of state-of-the-art commercial and academic algorithms. However, reproducing or analysing their results may be impossible due to the unattainability of data and/or tested systems.
\end{enumerate}

Following common findings for the evaluated biometric algorithms can be discerned:

\begin{description}
\item[Recognition] One result which appears to be mostly consistent across surveyed studies is that of worse biometric performance (both in terms of false positives and false negatives) for female subjects (see \eg~\cite{Grother-NIST-FRVTBias-2019,Klare-FaceDemographics-2012}). Furthermore, several studies associated race as a major factor influencing biometric performance. However, the results were not attributed to a specific race being inherently more challenging. Rather, the country of software development (and presumably the training data) appears to play a major role; in this context, evidence of the ``other-race'' effect in facial recognition has been found~\cite{Phillips-FaceRace-2011}, \eg algorithms developed in Asia were more easily recognising Asian individuals and conversely algorithms developed in Europe were found to be more easily recognising Caucasians. Finally, the age has been determined to be an important factor as well -- especially the very young subjects were a challenge (with effects of ageing also playing a major role). Grother~\etal~\cite{Grother-NIST-FRVTBias-2019} presented hitherto the largest and most comprehensive study of demographic bias in biometric recognition. Their benchmark showed that false-negative differentials usually vary by a factor of less than 3 across the benchmarked algorithms. On the other hand, the false-positive differentials were much more prevalent (albeit not universal) and often larger, \ie varying by two to three orders of magnitude across the benchmarked algorithms\footnote{Note that this is a very high-level summary to illustrate the general size of the demographic differentials. The experimental results are much more nuanced and complex, as well as dependent on a number of factors in the used data, experimental setup, and the algorithms themselves.}. Most existing studies considered biometric verification, with only a few addressing biometric identification. Estimating bias in biometric identification is non-trivial, due to the contents of the screening database being an additional variable factor susceptible to bias. Specifically, in addition to potential biases in the biometric algorithms themselves, certain biases stemming from data acquisition might occur and be propagated (\eg historical and societal biases having impact on the demographic composition of a criminal database). Consequently, demographic bias estimation in biometric identification is an interesting and important item for future research.
\item[Classification and estimation] Scientific literature predominantly studied face as the biometric characteristic, since the facial region contains rich information from which demographic attributes can be estimated. Several studies showed substantial impact of sex and race on the accuracy of demographic attribute classification. In particular, numerous commercial algorithms exhibited significantly lower accuracy \wrt dark-skinned female subjects (see \eg~\cite{Buolamwini-GenderShades-2018,Raji-ActionableAuditing-2019}). Research on classification of sex from iris and periocular images exists, but biases in those algorithms have not yet been studied. Additionally, it is not clear if such classifiers rely on actual anatomical properties or merely the application of mascara~\cite{Iris-Mascara-2017}.
\item[Quality assessment] Most existing studies conducted experiments using fingerprint-based biometrics. This could be partially caused by the standardisation of reliable fingerprint quality assessment metrics~\cite{ISO-FingerprintQuality-2017}, whereas this remains an open challenge for the face characteristic~\cite{Galbally-FaceIdentificationSchengen-2019}. The existing fingerprint quality assessment studies consistently indicated that the extreme ranges of the age distribution (infants and elderly) can pose a challenge for current systems~\cite{Galbally-FingerprintAge-2019}. Correlations between the quality metrics of facial images (obtained using state-of-the-art estimators) and demographic covariates were recently pointed out in a preliminary study~\cite{Terhorst-BiasFaceQuality-2020}. Additional non-obvious, hidden biases can also occur. For example, the presence of eyeglasses~\cite{Drozdowski-GlassesDetection-ICB-2018, OsorioRoig-GlassesDetectionVW-SITIS-2018} or contact lenses~\cite{Baker-IrisLenses-2010} lowers the sample quality and biometric performance under objective metrics in iris recognition systems. The demographics disproportionately afflicted with myopia (\ie most likely to wear corrective eyewear) are those from ``developed'' countries and East Asia~\cite{Dolgin-NatureNews-Myopia-2015}. Admittedly, the inability of the algorithms to compensate for the presence of corrective eyewear might be argued not to be a bias \textit{per se}. This argument notwithstanding, specific demographic groups could clearly be disadvantaged in this case -- either by increased error rates or the requirement for a more elaborate (especially for contact lenses) interaction with the acquisition device. Issues such as this one push the boundaries of what might be considered biased or fair in the context of biometric systems and constitute an interesting area of future technical and philosophical research.
\end{description}

In addition, it is necessary to point out potential issues in surveyed studies, such as:

\begin{itemize}
\item \emph{Differences in experimental setups}, used toolchains and datasets, training-testing data partitioning, imbalanced datasets \etc
\item Statistical significance of the results due to relatively \emph{small size of the used datasets} in most cases (except \eg~\cite{Grother-NIST-FRVTBias-2019,Galbally-FingerprintAge-2018}).
\item \emph{Lack} of a single \emph{definition of bias/fairness} (see also subsection~\ref{subsec:fairness}), as well as a standardised methodology and metrics for conducting evaluations.
\item Difficulty of sufficiently \emph{isolating the influence} of demographic factors from other important covariates (\eg pose and illumination).
\item Potential for \emph{bias propagation} from previous steps of the pipeline (\eg data acquisition).
\end{itemize}

Nevertheless, some results appear to be intuitive, \eg worse accuracies for women. These could be due to numerous reasons, such as: larger intra-class variations due to make-up~\cite{Rathgeb-FacialBeautification-2019}, occlusion by hairstyle and accessories, or pose differences due to women being shorter than men and cameras being calibrated with the height of men. Likewise, lower sample quality of infant fingerprints makes sense due to anatomical constraints and the fact that the size of the fingerprint area is considered as a relevant factor for fingerprint sample quality. In order to acquire high-quality fingerprint samples from very young data subjects, specialised hardware may be necessary (see \eg~\cite{Koda-InfantFingerprintSensor-2019}).

\subsection{Mitigation}
\label{subsec:mitigation}
Table~\ref{table:mitigationstudies} summarises the existing research in the area of bias mitigation in biometrics. Similarly to above, related work here focuses predominantly on face as biometric characteristic. In this context, mainly recognition and classification algorithms have been analysed. Generally speaking, the existing approaches can be assigned to following categories: 

\begin{description}
\item[Training] Learning-based methods have experienced a tremendous growth in accuracy and popularity in recent years. As such, the training step is of critical importance for the used systems and mitigation of demographic bias. The existing techniques mainly rely on demographically balanced training datasets (\eg~\cite{Wang-RacialFacesInTheWild-2019}) and synthetic data to enhance the training datasets (\eg~\cite{Kortylewski-BiasSyntheticData-2019}), as well as learning specialised loss or similarity functions (\eg~\cite{Terhorst-FaceBiasMitigation-2020}). A number of balanced training datasets has been released to the research community, as shown in table~\ref{table:datasets}.
\item[Dynamic selection] Deviating from preventing demographic bias, some methods attempted to employ a bias-aware approach. Examples in this category include dynamic selection of the recognition algorithms (\eg~\cite{Guo-AgeClassification-2010}) or decision thresholds (\eg~\cite{Krishnapriya-FaceRaceVariability-2019}) based on demographic attributes of the individual subjects.
\end{description}

In addition to the categories above, other approaches may be considered in the context of bias mitigation. For example, modelling of factors such as fingerprint growth can be used to improve the biometric recognition performance for children (see \eg~\cite{Preciozzi-FingerprintAge-2020}) and to mitigate the effects of ageing (see \eg~\cite{Haraksim-FingerprintMitigation-2019}). Other examples include de-identification and anonymisation methods (see \eg~\cite{Ribaric-Deidentfication-2017,Chhabra-Deidentification-2018}), whose primary use case is privacy-protection in biometrics. Such methods aim to remove, change, or obfuscate certain information (\eg demographics) either from the image (\eg~\cite{Mirjalili-GANPrivacy-2018}) or feature (\eg~\cite{Acien-ClassificationBias-2018,Morales-SensitiveNets-2019}) domain, often through a form of adversarial learning. One could hypothesise that a system trained on such data might not exhibit biases \wrt to the de-identified demographic covariates. However, the validity of such hypotheses has not yet been ascertained experimentally.

\begin{table}[!ht]
\centering
\caption{Summary of existing datasets for bias-related research in biometrics.}
\label{table:datasets}
\resizebox{\columnwidth}{!}{
\begin{tabular}{lllp{4cm}}
\toprule
\textbf{Reference} & \textbf{Characteristic} & \textbf{Size (images)} & \textbf{Details} \\
\midrule
Ricanek~\etal~\cite{Ricanek-MORPH-2006} & Face & 55.134 & Ageing research database with demographic labels. \\
Azzopardi~\etal~\cite{Azzopardi-GenderFERET-2016} & Face & 946 & Subset of FERET dataset balanced \wrt sex. \\
Buolamwini~\etal~\cite{Buolamwini-GenderShades-2018} & Face & 1.270 & Images of parliamentarians balanced \wrt sex and race. One image per subject, \ie not suitable for biometric recognition. \\
Alvi~\etal~\cite{Alvi-BiasRemoval-2018} & Face & 14.000 & Scraped images balanced \wrt race. \\
Alvi~\etal~\cite{Alvi-BiasRemoval-2018} & Face & 60.000 & Subset of IMDB dataset balanced \wrt sex and race. \\
Morales~\etal~\cite{Morales-SensitiveNets-2019} & Face & 139.677‬ & Subset of MegaFace dataset balanced \wrt sex and race. \\
Merler~\etal~\cite{Merler-DiF-2019} & Face & 964.873 & Demographic and geometric annotations for selected images from YFCC-100M dataset. \\
Hupont~\etal~\cite{Hupont-BiasFace-2019} & Face & 10.800 & Subset of CWF and VGG datasets balanced \wrt sex and race. \\
K{\"a}rkk{\"a}inen~\etal~\cite{Karkkainen-FairFace-2019} & Face & 108.501 & Subset of YFCC-100M dataset balanced \wrt sex, race, and age. \\
Wang~\etal~\cite{Wang-RacialFacesInTheWild-2019} & Face & 40.6070 & Subset of MS-Celeb-1M dataset balanced \wrt race. \\
Robinson~\etal~\cite{Robinson-FaceBias-2020} & Face & 20.000 & Subset of LFW dataset balanced \wrt sex and race. \\
Albiero~\etal~\cite{Albiero-FaceGender-2020} & Face & 42.134 & Subset of AFD dataset balanced \wrt sex. \\
\bottomrule
\end{tabular}
}
\end{table}

\section{Discussion}
\label{sec:discussion}
In this section, several issues relevant to the topic of this article are discussed. Concretely, subsection~\ref{subsec:fairness} addresses the topic of algorithmic fairness in general, while subsection~\ref{subsec:biasfairnessbiometrics} does so in the context of biometrics specifically. Subsection~\ref{subsec:socialimpact} illustrates the importance of further research on algorithmic bias and fairness in biometrics by describing the social impact of demographically biased systems.

\subsection{Algorithmic Fairness in General}
\label{subsec:fairness}
The challenge of fairness is common in machine learning and computer vision, \ie it is by no means limited to biometrics. A survey focusing on issues and challenges associated with algorithmic fairness was conducted among industry practitioners by Holstein~\etal~\cite{Holstein-FairnesML-2019}. For a comprehensive overview of bias in automated algorithms in general, the reader is referred to \eg~\cite{Mehrabi-BiasSurvey-2019,Du-Fairness-2019}. In addition to algorithmic fairness, algorithmic transparency, explainability, interpretability, and accountability (see \eg~\cite{Diakopoulos-AccountabilityShort-2016,Kroll-AccountableAlgorithms-2016,Ribeiro-ExplainingPredictions-ICKDDM-2016,New-AlgorithmicAccountability-2018}) have also been heavily researched in recent years both from the technical and social perspective. The current research in the area of algorithmic fairness concentrates on the following topics:

\begin{itemize}
\item Theoretical and formal definitions of bias and fairness (see \eg~\cite{Mehrabi-BiasSurvey-2019,Verma-FairnessDefinitions-2018,Hutchinson-FairnessSurvey-2019}).
\item Fairness metrics, software, and benchmarks (see \eg~\cite{Hardt-EqualityLearning-2016,Saleiro-BiasToolkit-2018,Bellamy-AIFairness-2018}).
\item Societal, ethical, and legal aspects of algorithmic decision-making and fairness therein (see \eg~\cite{Pasquale-BlackBoxSociety-2015,Tufekci-AlgorithmicHarms-2015,Corbett-AlgorithmicFairness-2017,FRA-BigData-2018,FRA-BiasFundamentalRights-2019}).
\item Estimation and mitigation of bias in algorithms and datasets (see \eg~\cite{Torralba-DatasetBias-2011,Zemel-FairRepresentations-2013,Shaikh-MLFairness-2017,Fernandez-ImbalancedLearning-2018,Zhang-BiasMitigation-2018,Roy-InformationLeakage-2019}).
\end{itemize}

Despite decades of research, there exists no single agreed coherent definition of algorithmic fairness. In fact, dozens of formal definitions (see \eg~\cite{Verma-FairnessDefinitions-2018,Hutchinson-FairnessSurvey-2019}) have been proposed to address different situations and possible criteria of fairness\footnote{See also \url{https://towardsdatascience.com/a-tutorial-on-fairness-in-machine-learning-3ff8ba1040cb} and \url{https://fairmlbook.org/} for visual tutorials on bias and fairness in machine learning.}. Certain definitions, which are commonly used and advocated for, are even provably mutually exclusive~\cite{Friedler-FairnessImpossibility-2016}.
Therefore, depending on the definition of fairness one chooses to adopt, a system can effectively always be shown to exhibit some form of bias. As such, the ``correct'' approach is essentially application-dependent. This in turn necessitates a keen domain knowledge and awareness of those issues from the system operators and stakeholders, as they need to select the definitions and metrics of fairness relevant to their particular use case. Research in this area strongly suggests that the notion of fairness in machine learning is context-sensitive~\cite{Green-FairnessContext-2018,Liu-FairnessContext-2018}; this presumably also applies to the field of biometrics, especially for the machine learning-based systems. In the next subsection, the notions of fairness and bias are discussed in the context of biometrics specifically based on the literature surveyed in section~\ref{sec:biasinbiometrics}.

\subsection{Algorithmic Fairness in Biometrics}
\label{subsec:biasfairnessbiometrics}
Although the topic of demographic bias and fairness in biometrics has emerged relatively recently, it has quickly established itself as an important and popular research area. Several high-ranking conferences featured special sessions\footnote{\url{https://sites.google.com/view/wacv2020demographics}}\textsuperscript{,}\footnote{\url{https://sites.google.com/site/eccvbefa2018}}\textsuperscript{,}\footnote{\url{https://dasec.h-da.de/wp-content/uploads/2020/01/EUSIPCO2020-ss_bias_in_biometrics.pdf}}, NIST conducted large-scale evaluations~\cite{Grother-NIST-FRVTBias-2019}, while ISO/IEC is currently preparing a technical report on this subject~\cite{ISO-Bias}. Likewise, a significant number of scientific publications has appeared on this topic (surveyed in section~\ref{sec:biasinbiometrics}). Existing studies concentrated on face-based biometrics -- more research is urgently needed for other biometric characteristics, \eg fingerprints~\cite{Marasco-BiasFingerprint-2019}.

Existing studies primarily address the following aspects: 

\begin{enumerate}
\item Evaluations with the aim of quantitatively ascertaining the degree of demographic bias in various biometric algorithms.
\item Methods which seek to mitigate the effects of demographic bias in various biometric algorithms.
\end{enumerate}

Existing bias estimation studies have uncovered new trends \wrt algorithmic bias and fairness in biometric algorithms (recall subsection~\ref{subsec:estimation}). However, it should be noted, that:

\begin{enumerate}
\item In many cases the biases were algorithm-specific, \ie while given the same benchmark-dataset some algorithms exhibited a bias (\eg lower biometric performance for a certain demographic group), others did not. In aggregate, however, the existing studies did seem to agree on certain points, as described in subsection~\ref{subsec:estimation}.
\item While a high relative increase in error rates for a certain demographic group may appear quite substantial, its importance in absolute terms could be negligible, especially for very accurate algorithms which hardly make any errors whatsoever~\cite{Grother-NIST-FRVTBias-2019}.
\end{enumerate}

Those caveats notwithstanding, the commitment of the academic researchers and commercial vendors to researching algorithmic fairness is especially important for the public perception of biometric technologies. The field of algorithmic fairness in the context of biometrics is in its infancy and a large number of research areas are yet to be comprehensively addressed (\cf subsection~\ref{subsec:fairness}):

\begin{enumerate}
\item Limited theoretical work has been conducted in this field specifically focusing on biometrics. Indeed, the majority of publications surveyed in section~\ref{sec:biasinbiometrics} do not approach the notions of bias and fairness rigorously; rather, they tend to concentrate on an equivalent of some of the simpler statistical definitions, such as group fairness and error rate parity. Extending the existing estimation and mitigation works, for example to consider other and more complex notions of fairness (see \eg~\cite{Terhorst-BiasMitigation-2020}) could be seen as important future work in the field. Likewise, investigating trade-offs between biometric performance, fairness, user experience, social perceptions, monetary costs, and other aspects of the biometric systems might be of interest.
\item In addition to empiric studies (especially in the case of bias mitigation, see subsection~\ref{subsec:mitigation}), stricter theoretical approaches need to be pursued in order to provably demonstrate the bias-mitigating properties of the proposed methods. 
\item Isolating the effects of the demographic factors from other confounding factors (\ie the environmental and subject-specific covariates, such as illumination and use of accessories) is a challenging task, which is not sufficiently addressed in many existing studies. An example of a study which partially addressed those issues in a systematic manner is the work of Grother~\etal~\cite{Grother-NIST-FRVTBias-2019}.
\item More complex analyses based on demographic attributes and combinations thereof (intersectionality) could be conducted for a more detailed and nuanced view of demographic biases in biometric systems.
\item Comprehensive independent benchmarks utilising various algorithmic fairness measurement methodologies and metrics are, as of yet, lacking. Only recently, in~\cite{Grother-NIST-FRVTBias-2019}, first independent benchmarks of biometric recognition algorithms have been conducted. Similar and more extensive benchmarks for other biometric algorithms (recall subsection~\ref{subsec:algorithms}) are needed.
\item Large-scale datasets designed specifically for bias-related research need to be collected. The existing datasets only pertain to face-based biometrics (see table~\ref{table:datasets}).
\item Humans are known to exhibit a broad range of biases~\cite{Evans-HumanBias-1989,Parasuraman-AutomationBias-2010}. The influence of those factors on biometric algorithm design, interactions with and use of biometric systems, as well as perceptions of biometric systems could be investigated.
\end{enumerate}

In the next subsection, the possible consequences of failing to appropriately address the issues of algorithmic fairness in biometrics are discussed.

\subsection{Social Impact}
\label{subsec:socialimpact}
Numerous studies described the potential of real harms as a consequence of biased algorithmic decision-making systems~\cite{Tufekci-AlgorithmicHarms-2015,Kirkpatrick-AlgorithmicBias-2016} in general. Regarding biometric systems in particular, facial recognition technologies have been the main focus of such discussions (see \eg~\cite{Hallowell-FaceRecognitionEthics-2019}). Concering the notions of bias and fairness, in addition to being context-sensitive (recall subsection~\ref{subsec:fairness}), one might argue the impact assessments to also be purpose-sensitive. Specifically, depending on application scenario, the impact and importance of systemic biases might differ significantly. As an example, consider an application of biometrics in cooperative access control systems or personal devices. A demographic bias in such a system might cause a certain demographic group to be inconvenienced through additional authentication attempt(s) being necessary due to false negative errors. On the other hand, the stakes are much higher in \eg a state surveillance scenario. There, demographic biases could directly cause substantial personal harms, \eg higher (unjustified) arrest rates~\cite{Garvie-PerpetualLineUp-2016}, due to false positive errors.

At the same time, it is also clear that biometric recognition technology can be highly accurate. Taking the recently contested facial recognition as an example, given prerequisites such as a high-resolution camera, proper lighting and image quality controls, as well as high-quality comparison algorithms, the absolute error rates can become vanishingly small~\cite{Grother-NIST-FRVTBias-2019}, thereby potentially rendering the relative imbalance of error rates across demographic groups insignificant. 

It should be noted that there are no indications of the algorithmic biases in biometrics being deliberately put into the algorithms by design; rather, they are typically a result of the used training data and other factors. In any case, one should also be mindful, that as any technology, biometrics could be used in malicious or dystopian ways (\eg privacy violations through mass-surveillance~\cite{Kindt-BiometricsPrivacy-2016} or ``crime prediction''~\cite{Wu-CriminalInference-2016}). Consequently, a framework for human impact assessments~\cite{Calvo-ImpactAssessments-2020} should be developed for biometrics as soon as possible. A pro-active and cognizant approach could foster awareness among the citizens and policymakers, as well as contribute to minimising potential negative perception of biometric technology and innovation by individuals and society as a whole.

In a broader context, algorithmic bias and fairness is one of the topics in the larger discourse on ethical design in artificial intelligence (AI) systems~\cite{Bryson-EthicalDesignStandardising-2017}, most prominently encompassing:

\begin{itemize}
\item Transparency,
\item Accountability,
\item Explainability, and
\item Fairness.
\end{itemize}

Currently, the legal and societal scrutiny of the technologies utilising automated decision systems seems to be insufficient. However, recent legislation in the European Union~\cite{EU-GDPR-2016,Goodman-EUAlgorithmicDecisionRights-2016} constitutes a step in the that direction. Below, several social and technological provisions, which might be considered in this context, are listed.

\begin{itemize}
\item Carefully \emph{selecting the data used to train} the algorithms is the first and perhaps the most important step: inherent biases in training data should be avoided wherever possible. Furthermore, the size of the dataset matters -- some systems have been reported to be trained on very small datasets (in the order of thousands of items), which is usually wholly insufficient to show that an approach generalises well.
\item Higher degree of \emph{transparency} and/or independent insight into data and algorithms, as well as validation of the results could be established to foster the public trust and acceptance of the systems.
\item Thresholds for \emph{acceptable accuracy} (\ie how much the systems can err) could be established legally (potentially in a system purpose-sensitive manner), as well as reviewed and validated periodically.
\item Special \emph{training of the systems' personnel} could be established to make them aware of the potential issues and to establish proper protocols for dealing with them.
\item \emph{Due diligence} could be legally expected from vendors of such systems, \ie in reasonably ensuring some or all of aforementioned matters and rectifying problems as they come up. Additionally, certain accountability provisions could be incorporated to further facilitate this.
\end{itemize}

The issues of fairness (including algorithmic fairness) are complicated from the point of view of the legislation -- a somewhat deep understanding of statistics, formal fairness definitions, and other concepts is essential for an informed discourse. Furthermore, the ethical and moral perceptions and decisions are not uniform across different population demographics and by geographical location (see \eg Awad~\etal~\cite{Awad-MoralMachine-2018}). This reinforces an important dilemma regarding the regulation of automated decision systems -- since many situations are morally and ethically ambiguous to humans, how should they be able to encode ethical decision-making into laws? Once that issue is somehow surmounted, there also remains the issue of feasibility of technical solutions, as described in the previous two subsections.

Currently, many laws and rules exist (international treaties, constitutions of many countries, and employment law) which aim to protect against generic discrimination on the basis of demographics~\cite{EU-AntiDiscrimination-2018}. However, historically, the enforcement of those has been fraught with difficulties and controversies. In this context, the algorithmic decision systems are merely one of the most recent and technologically advanced cases. The policymakers and other stakeholders will have to tackle it in the upcoming years in order to develop a legal framework similar to those already governing other areas and aspects of the society~\cite{Nature-Editorial-2016}.
\vspace{-0.3cm}
\section{Summary}
\label{sec:summary}
This article has investigated the challenge of demographic bias in biometric systems. Following an overview of the topic and challenges associated therewith, a comprehensive survey of the literature on bias estimation and mitigation in biometric algorithms has been conducted. It has been found that demographic factors can have a large influence on various biometric algorithms and that current algorithms tend to exhibit some degree of bias \wrt certain demographic groups. Most effects are algorithm-dependent, but some consistent trends do also appear (as discussed in subsection~\ref{subsec:estimation}). Specifically, many studies point to a lower biometric performance for females and youngest subjects in biometric recognition systems, as well as lower classification accuracy for dark-skinned females in classification of demographic attributes from facial images. It should be noted that many of the studies conducted their experiments using relatively small datasets, which emphasises the need for large-scale studies. In general, a broad spectrum of open technical (and other) challenges exists in this field (see section~\ref{sec:discussion}).

Biased automated decision systems can be detrimental to their users, with issues ranging from simple inconveniences, through disadvantages, to lasting serious harms. This relevance notwithstanding, the topic of algorithmic fairness is still relatively new, with many unexplored areas and few legal and practical provisions in existence. Recently, a growing academic and media coverage has emerged, where the overwhelming consensus appears to be that such systems need to be properly assessed (\eg through independent benchmarks), compelled to some degree of transparency, accountability, and explainability in addition to guaranteeing some fairness definitions. Furthermore, it appears that, in certain cases, legal provisions might need to be introduced to regulate these technologies.

Automatic decision systems (including biometrics) are experiencing a rapid technological progress, thus simultaneously holding a potential of beneficial and harmful applications, as well as unintentional discrimination. Zweig~\etal~\cite{Zweig-AlgoritmsChances-2018} even argued that the issues (including, but not limited to bias and fairness) concerning algorithmic decision systems are directly related to the so-called ``quality of democracy'' measure of countries. As such, developing proper frameworks and rules for such technologies is a large challenge which the policymakers and the society as a whole must face in the upcoming future~\cite{Citron-TechnologicalDueProcess-WULR-2007,Citron-ScoredSociety-2014}.

\section*{Acknowledgements}
\label{sec:acknowledgements}
This research work has been funded by the German Federal Ministry of Education and Research and the Hessen State Ministry for Higher Education, Research and the Arts within their joint support of the National Research Center for Applied Cybersecurity ATHENE. A. Dantcheva was funded by the French Government (National Research Agency, ANR), under Grant ANR-17-CE39-0002.

\bibliographystyle{IEEEtran}
\bibliography{references}

\newpage

\begin{IEEEbiography}[{\includegraphics[width=1in,height=1.25in,clip,keepaspectratio]{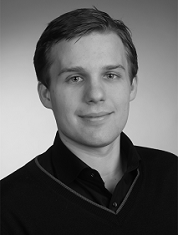}}]{Pawel Drozdowski}
is a Researcher with the Faculty of Computer Science, Hochschule Darmstadt (HDA), Germany. He pursues a Ph.D. degree at the Norwegian University of Science and Technology (NTNU). He co-authored over 15 technical publications in the field of biometrics. He won the Best Student Paper Runner-Up Award (WIFS'18) and Best Poster Award (BIOSIG'19). He represents the German Institute for Standardization (DIN) in ISO/IEC SC37 JTC1 SC37 on biometrics.
\end{IEEEbiography}

\vspace{-0.75cm}

\begin{IEEEbiography}[{\includegraphics[width=1in,height=1.25in,clip,keepaspectratio]{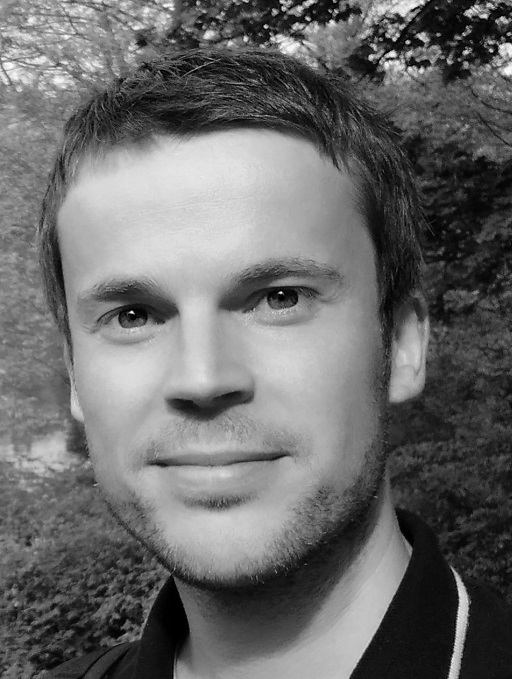}}]{Dr. Christian Rathgeb}
is a Senior Researcher with the Faculty of Computer Science, Hochschule Darmstadt (HDA), Germany. He is a Principal Investigator in the National Research Center for Applied Cybersecurity ATHENE. His research includes pattern recognition, iris and face recognition, security aspects of biometric systems, secure process design and privacy enhancing technologies for biometric systems. He co-authored over 100 technical publications in the field of biometrics. He is a winner of the EAB - European Biometrics Research Award 2012, the Austrian Award of Excellence 2012, Best Poster Paper Awards (IJCB'11, IJCB'14, ICB'15) and the Best Paper Award Bronze (ICB'18).
\end{IEEEbiography}

\vspace{-0.75cm}

\begin{IEEEbiography}[{\includegraphics[width=1in,height=1.25in,clip,keepaspectratio]{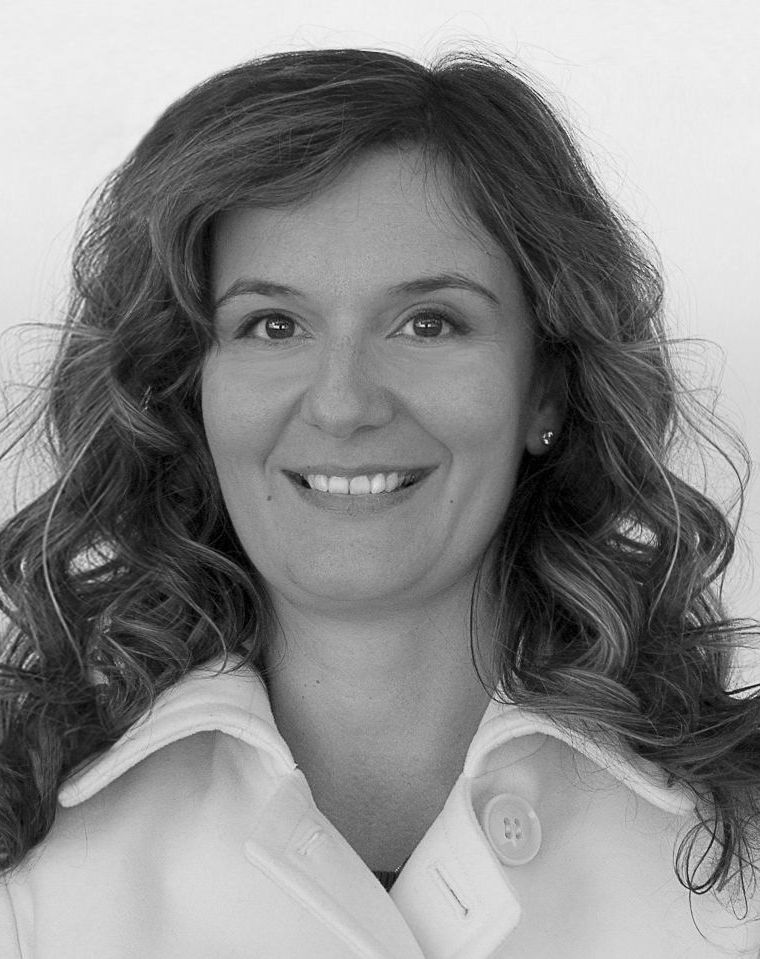}}]{Dr. Antitza Dantcheva} 
is a Research Scientist (CRCN) with the STARS team of INRIA Sophia Antipolis, France. Her research interests include computer vision and face analysis, where she has been working in appearance and dynamic analysis for healthcare and security. She has received the French National Research Agency (ANR) JCJC Young Researcher Grant in 2017. She was a recipient of the Best Presentation Award (ICME'11), the Best Poster Award (ICB'13), Tabula Rasa Spoofing Award in 2013, Best Paper Award (Runner Up) (ISBA'17), as well as the Best Poster Award (FG'19). She was in the winning team of the ECCV 2018 Challenge on Bias Estimation in Face Analysis.
\end{IEEEbiography}

\vspace{-0.75cm}

\begin{IEEEbiography}[{\includegraphics[width=1in,height=1.25in,clip,keepaspectratio]{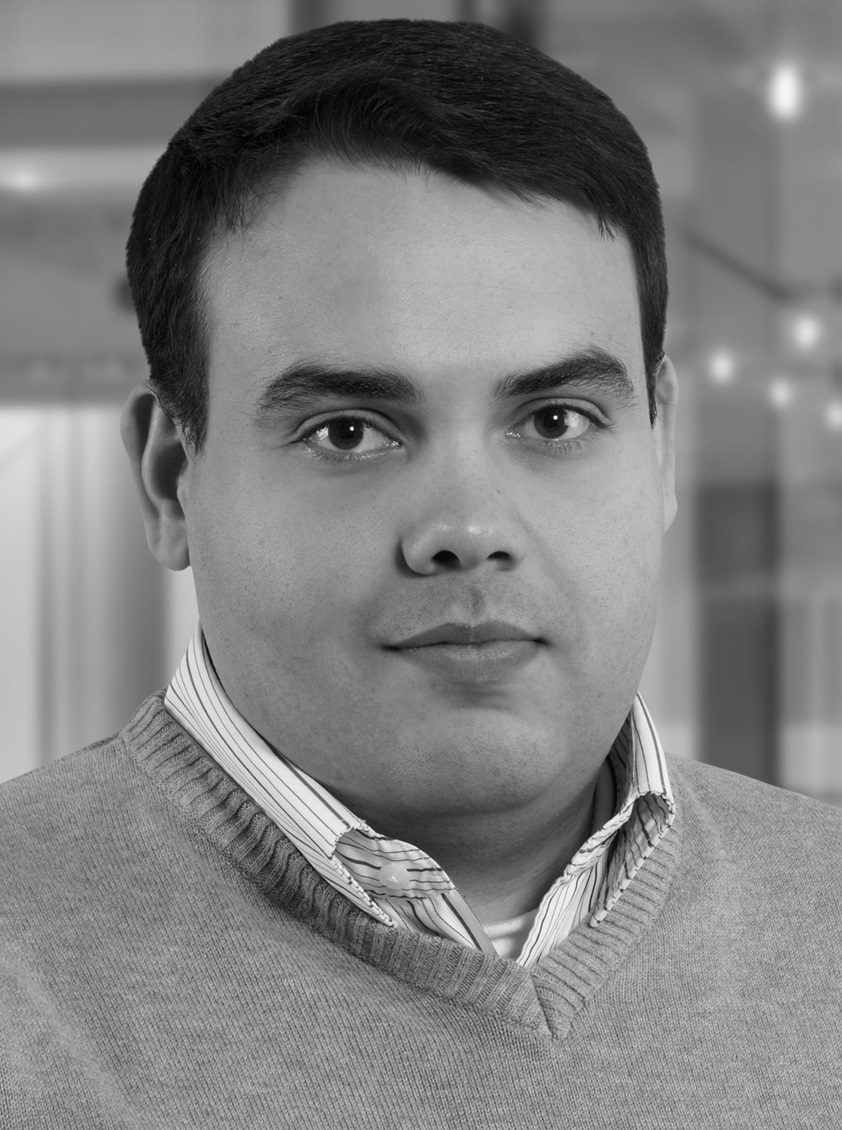}}]{Dr. Naser Damer}
is a senior researcher at Fraunhofer IGD. He received his PhD in computer science from the Technical University of Darmstadt (2018). His main research interests lie in the fields of biometrics, machine learning and information fusion. He published more than 60 scientific papers in these fields. Dr. Damer is a Principal Investigator at the National Research Center for Applied Cybersecurity ATHENE in Darmstadt, Germany. He serves as an associate editor for the Visual Computer journal and represents the German Institute for Standardization (DIN) in ISO/IEC SC37 biometrics standardization committee.
\end{IEEEbiography}

\vspace{-0.75cm}

\begin{IEEEbiography}[{\includegraphics[width=1in,height=1.25in,clip,keepaspectratio]{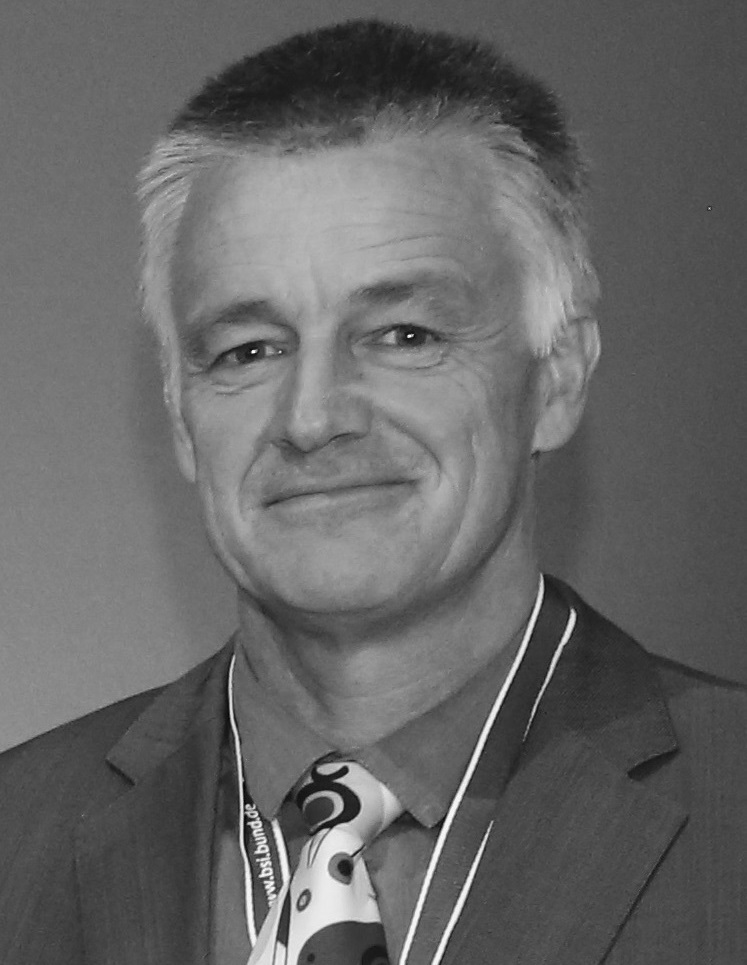}}]{Prof. Dr. Christoph Busch}
is member of the Norwegian University of Science and Technology (NTNU). He holds a joint appointment with Hochschule Darmstadt (HDA), Germany. Further he lectures Biometric Systems at Technical University of Denmark (DTU) since 2007. Christoph Busch co-authored more than 500 technical papers and has been a speaker at international conferences. He served for several conferences and journals as reviewer (e.g. ACM-SIGGRAPH, ACM-TISSEC, IEEE CG\&A, IEEE PAMI). He is also an appointed member of the editorial board of the IET journal on Biometrics and of IEEE TIFS journal. Furthermore, he chairs the German biometrics working group and is board member of the European Association for Biometrics (EAB). He chairs the German standardization body on Biometrics (DIN -- NIA37) and is convenor of WG3 in ISO/IEC JTC1 SC37 on Biometrics.
\end{IEEEbiography}

\end{document}